\documentclass{webofc}
\usepackage[varg]{txfonts}   
\usepackage{slashed}
\begin{document}
\title{Early deconfinement of asymptotically conformal color-superconducting quark matter in neutron stars}
%
%

\author{
\firstname{Oleksii} \lastname{Ivanytskyi}\inst{1}\fnsep\thanks{\email{oleksii.ivanytskyi@uwr.edu.pl}} \and
\firstname{David} \lastname{Blaschke}\inst{1}\fnsep\thanks{\email{david.blaschke@uwr.edu.pl}} \and
\firstname{Tobias} \lastname{Fischer}\inst{1}\fnsep\thanks{\email{tobias.fischer@uwr.edu.pl}} \and
\firstname{Andreas} \lastname{Bauswein}\inst{2}\fnsep\thanks{\email{a.bauswein@gsi.de }}
}

\institute{Institute of Theoretical Physics, University of Wroclaw, Max Born Place 9, 50-204 Wroclaw, Poland
\and
GSI Helmholtzzentrum f\"ur Schwerionenforschung, Planckstra{\ss}e 1, 64291 Darmstadt, Germany}

\abstract{
We present a relativistic density functional approach to color superconducting quark matter that mimics quark confinement by a fast growth of the quasiparticle selfenergy in the confining region.
The approach is shown to be equivalent to a chiral model of quark matter with medium dependent couplings.
While the (pseudo)scalar sector of the model is fitted to the vacuum phenomenology of quantum chromodynamics, the strength of interaction in the vector and diquark channels is varied in order to provide the best agreement with the observational constraints on the mass-radius relation and tidal deformability of neutron stars modelled with our approach. 
In order to recover the conformal behavior of quark matter at asymptotically high densities we introduce a medium dependence of the vector and diquark couplings motivated by the non-perturbative gluon exchange.
Our analysis signals that the onset of deconfinement to color superconducting quark matter is likely to occur in neutron stars with masses below $1.0~{\rm M}_\odot$.
}
\maketitle

\section{Introduction}
\label{intro}

The high density region of the phase diagram of quantum chromodynamics (QCD) can be probed only with the multi-messenger observations of neutron stars (NS) in isolation, binary mergers and supernova explosions \cite{Bauswein:2022vtq}.
Therefore, recent measurements of the radii of a $1.4~{\rm M}_\odot$ NS $R_{1.4}=11.7^{+0.86}_{-0.81}$ km \cite{Dietrich:2020efo} and of a $2.0~{\rm M}_\odot$ NS $R_{2.0}=13.7^{+2.6}_{-1.5}$ km
\cite{Miller:2021qha} set important constraints on the equation of state (EoS) of dense matter.
Supplemented with the limitation on the tidal deformability of a $1.4~{\rm M}_\odot$ NS $\Lambda_{1.4}=190^{+390}_{-120}$ extracted from the gravitational wave signal of the NS merger event GW170817 \cite{LIGOScientific:2018cki} these new measurements of the NS radii pose a challenge for purely hadronic EoS which now are not quite excluded, but become only marginally consistent with these data. 
In particular because of the appearance of hyperons and heavier (multi-)baryon states at NS masses above $\sim 1.4~{\rm M}_\odot$, and their effect to soften the EoS and thus to lower the maximum mass and the radii of NS which leads to the 
"hyperon puzzle" there is tension between realistic, purely hadronic EoS and NS mass and radius measurements. 
In principle, this softening can be compensated by introducing additional repulsion.
This, however, makes hadronic EoSs inconsistent within the above constraint on $\Lambda_{1.4}$.
An elegant way out of this dilemma is provided by an EoS with a low onset density for the transition to stiff quark matter in the NS core so that both of the above radius conditions and the limitation on tidal deformability can be fulfilled simultaneously. 
It is worth mentioning that early deconfinement in cold quark-hadron matter is also supported by studies of the multi-messenger observational data on the NS masses and radii \cite{Blaschke:2020vuy} and by a systematic Bayesian analysis of these data \cite{Shahrbaf:2021cjz}.

As it has been discussed in detail in the review by Baym et al. \cite{Baym:2017whm}, the scenario with an early onset of quark matter requires a quark matter EoS with a repulsive vector mean field for the stiffening and strong color superconductivity with sufficiently large diquark pairing gap for the early onset of the deconfinement transition. 
Recently, an EoS agreeing with these requirements has been worked out within the approach of relativistic density functional (RDF), providing an efficient phenomenological confinement of quarks at low temperatures and densities \cite{Ivanytskyi:2022bjc}.
An important conclusion drawn from the modelling with this approach to the QCD phase diagram is that the temperature at the quark-to-hadron transition grows along the adiabates \cite{Ivanytskyi:2022wln} driving the trajectories of evolution of proto NS produced in the supernova explosions toward the phase diagram regions accessed in the NS mergers and experiments with heavy ion collisions.
In this contribution we present a further development of the confining RDF approach to color superconducting quark matter, which provides its conformal behavior at asymptotically high densities \cite{Ivanytskyi:2022bjc}.
This extends the range where the present approach is consistent with the recent constraint from perturbative QCD (pQCD) \cite{Komoltsev:2021jzg} that is able to discredit many EoSs of the NS matter.

\section{Relativistic density functional for quark matter}
\label{sec1}

The RDF approach for two flavors of quarks with degenerate mass $m$ is given by the Lagrangian written in terms of the field $q^T=(u~d)$ as \cite{Ivanytskyi:2022oxv}
\begin{eqnarray}
\label{I}
\mathcal{L}=\overline{q}(i\slashed\partial- m)q-
\mathcal{L}_V+\mathcal{L}_D-\mathcal{U}.
\end{eqnarray}
Vector repulsion and diquark pairing channels enter $\mathcal{L}$ through
\begin{eqnarray}
\label{II}
\mathcal{L}_V&=&G_V(\overline{q}\gamma_\mu q)^2,\\
\mathcal{L}_D&=&G_D(\overline{q}i\gamma_5\tau_2\lambda_A q^c)(\overline{q}^ci\gamma_5\tau_2\lambda_A q)
\end{eqnarray}
with $G_V$ and $G_D$ being the corresponding coupling constants and summation is performed over the color index $A=2,5,7$. The chirally symmetric potential $\mathcal{U}$ is inspired by the string-flip model (SFM) \cite{Kaltenborn:2017hus} and represents scalar and pseudoscalar interaction channels. 
It is controlled by the constant parameters $\alpha$ and $D_0$. Thus
\begin{eqnarray}
\mathcal{U}&=&D_0\left[(1+\alpha)\langle \overline{q}q\rangle_0^2
-(\overline{q}q)^2-(\overline{q}i\gamma_5\vec\tau q)^2\right]^{\frac{1}{3}}\nonumber\\
\label{IV}
&\simeq&\mathcal{U}_{\rm MF}+
(\overline{q}q-\langle\overline{q}q\rangle)\Sigma_{\rm MF}-
G_{S}(\overline{q}q-\langle\overline{q}q\rangle)^2-
G_{PS}(\overline{q}i\gamma_5\vec\tau q)^2,
\end{eqnarray}
where $\langle \overline{q}q\rangle_0$ is the chiral condensate in the vacuum. 
In what follows the subscript index ``0'' labels the quantities defined in the vacuum.
The next step is to expand this potential up to second order around the mean-field solutions $\langle \overline{q}q\rangle$ and $\langle \overline{q}i\gamma_5\vec\tau q\rangle=0$. 
The subscript index ``$\rm MF$'' labels quantities defined at mean field.
Eq. (\ref{IV}) includes the mean-field scalar self-energy of quarks $\Sigma_{\rm MF}=\partial\mathcal{U}_{\rm MF}/\partial\langle\overline{q}q\rangle$ and the medium dependent scalar $G_S=-\partial\mathcal{U}_{\rm MF}^2/\partial\langle\overline{q}q\rangle^2/2$ and pseudoscalar $G_{PS}=-\partial\mathcal{U}_{\rm MF}^2/\partial\langle\overline{q}i\gamma_5\vec\tau q\rangle^2/6$ couplings. 

\begin{figure}[t]
\centering
\includegraphics[width=0.32\columnwidth]{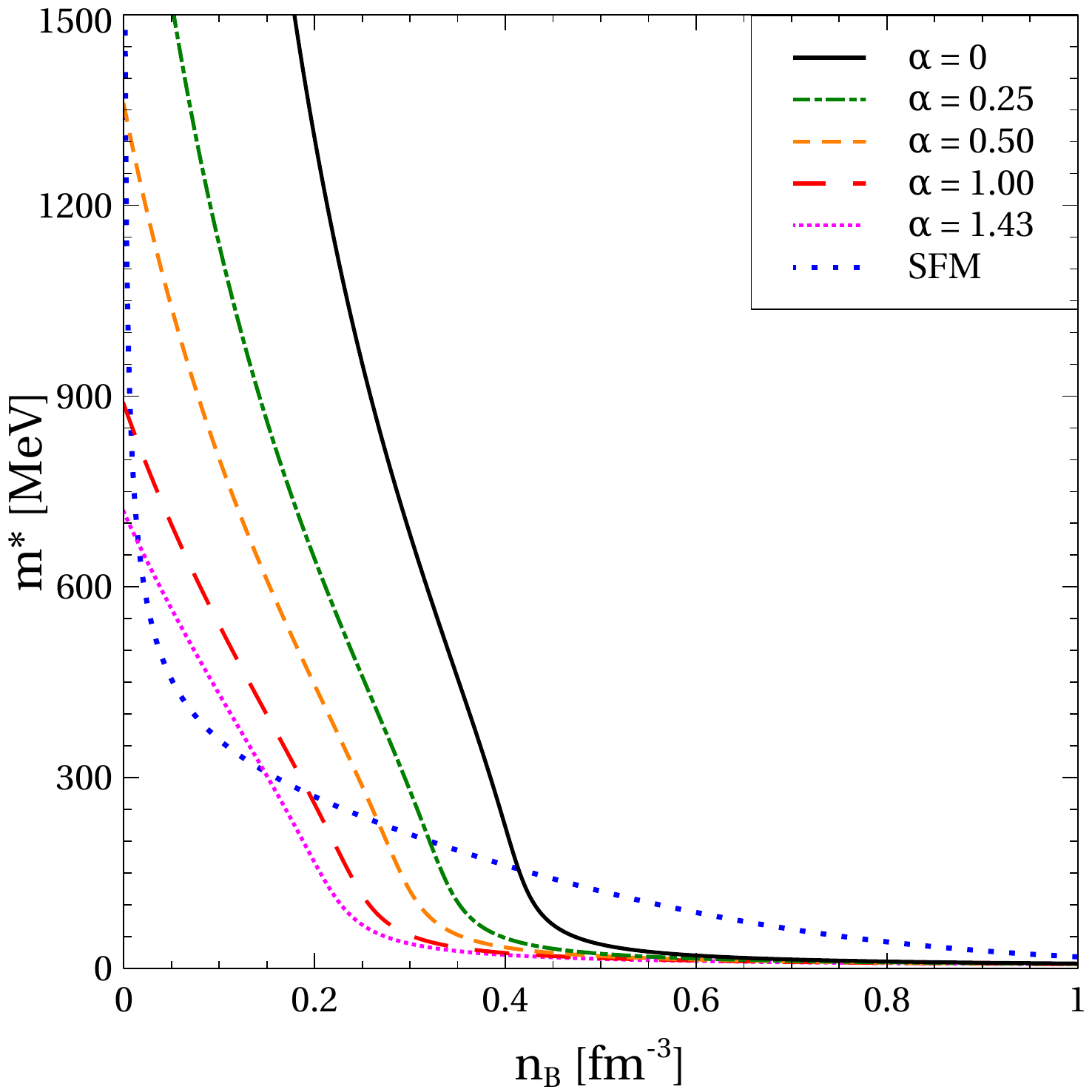}
\includegraphics[width=0.32\columnwidth]{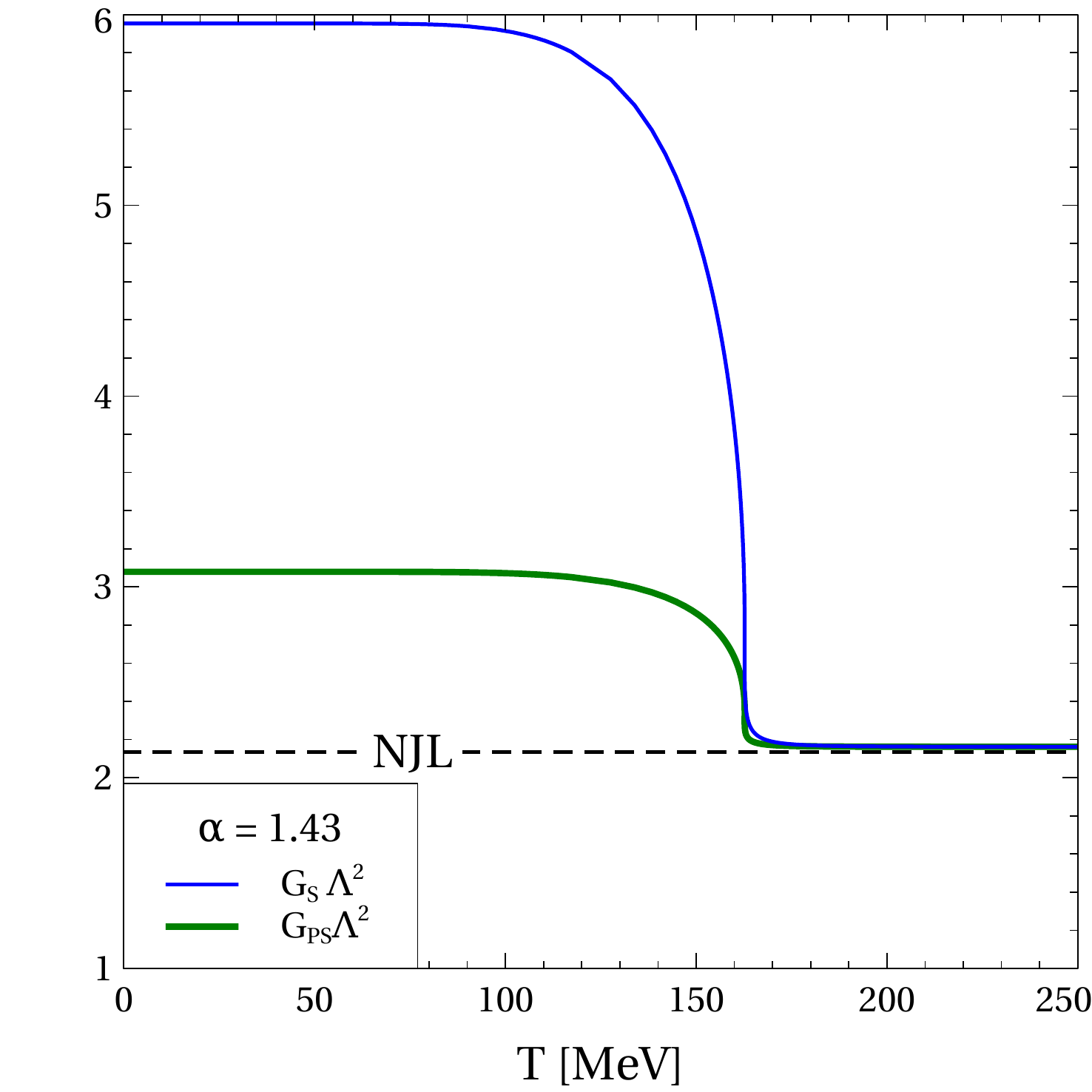}
\includegraphics[width=0.32\columnwidth]{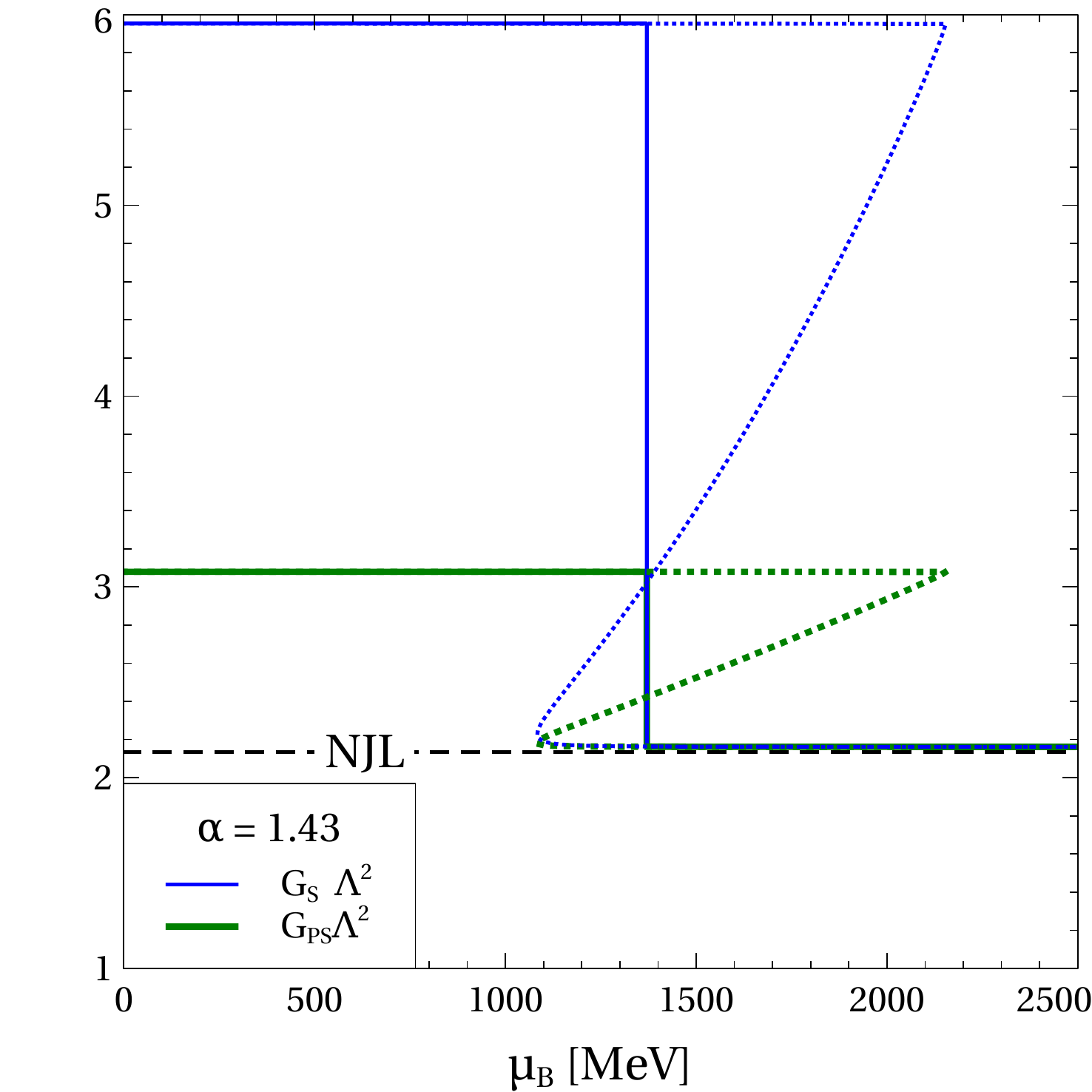}
\caption{Effective quark mass $m^*$ as function of baryon density $n_B$ (left panel), scaled effective scalar $G_S\Lambda^2$ and pseudoscalar $G_{PS}\Lambda^2$ couplings as functions of temperature $T$ at $\mu_B=0$ (middle panel) and baryonic chemical potential $\mu_B$ at $T=0$ (right panel).
The blue dotted line on the left panel is obtained within the SFM with $\alpha_{SFM}=0.39~{\rm fm}^{-3}$ \cite{Kaltenborn:2017hus}.
Dashed lines on the middle and right panels represent the NJL value $G\Lambda^2=2.14$  \cite{Ratti:2005jh}. Dotted curves on the middle panel indicate the unstable parts that are removed by applying the Maxwell construction. Calculations are performed for symmetric quark matter, $G_V=G_D=0$, $\alpha$ specified in the legend and the rest of the model parameters from Table \ref{table1}.}
\label{fig1}
\end{figure}

The expansion of the potential $\mathcal{U}$ brings the present RDF approach to the form of a NJL model. 
This feature was used in Ref. \cite{Ivanytskyi:2022oxv} in order to apply the common strategy of fitting the model parameters to vacuum phenomenology of QCD. 
The most important in this context are the pion mass $M_\pi$ and decay constant $F_\pi$, the scalar meson mass $M_\sigma$ and the vacuum value of the chiral condensate per flavor $\langle\overline{l}l\rangle_0$.
Note, the value of $M_\sigma$ was chosen in agreement with the mass of the narrowest candidate for the scalar meson role \cite{PhysRevD.98.030001}.
These physical quantities along with the resulting model parameters are listed in Table \ref{table1}.
One of the parameters not discussed before is the range $\Lambda$ of the smooth three-momentum cut-off introduced via a Gaussian formfactor in order to regulate divergent zero-point terms.
The value of $\langle\overline{l}l\rangle_0$ is somewhat larger than the ones obtained from
QCD sum rules at the renormalization scale of 1 GeV \cite{Jamin:2002ev}, which is a common situation for the NJL type chiral quark models \cite{Grigorian:2006qe}. 
This parameterization yields the pseudocritical temperature at $\mu_B=0$ defined by the peak of the chiral susceptibility $T_\chi=163$ MeV. The vacuum observables mentioned above are not affected by the values of the vector and diquark couplings.
Therefore, they are treated as free parameters parameterized by the dimensionless quantities $\eta_V=G_V/G_{S0}$ and $\eta_D=G_D/G_{S0}$. The pair of numbers $(\eta_V,\eta_D)$ is used in order to label the model EoSs considered below. 
We also require $\eta_D<0.78$ in order to prevent the formation of a color-superconducting state already in the vacuum \cite{Ivanytskyi:2022oxv}.
This maximal value of $\eta_D$ is independent of the vector coupling since the baryon density vanishes in the vacuum.

\begin{table}[h]
\centering
\caption{Parameters of the model used and resulting physical quantities}
\label{table1}       
\begin{tabular}{cccccccc}
\hline
$m$ & $\Lambda$ & $\alpha$ & $D_0\Lambda^{-2}$ &
$M_\pi$ & $F_\pi$ & $M_\sigma$ & $|\langle\overline{l}l\rangle_0|^{1/3}$\\\hline
4.2 MeV  & 573 MeV &    1.43  &  1.39 &
 140 MeV & 92 MeV &  980 MeV & 267 MeV\\\hline
\end{tabular}
\end{table}

The present approach provides efficient phenomenological confinement of quarks at small densities and temperatures by suppressing them due to large effective mass 
$m^*=m+\Sigma_{\rm MF}$.
In the vacuum $m^*_0=m-2/3~D_0\alpha^{-2/3}\langle\overline{q}q\rangle_0^{-1/3}$ is controlled by the value of $\alpha$.
For the chosen set of model parameters $m_0^*=718$ MeV.
It is remarkable that contrary to the (P)NJL model in the chiral limit the present one violates the BCS relation between the  chiral symmetry breaking mass gap in the vacuum and the critical temperature for its restoration.
The behavior of $m_0^*$, $G_S$ and $G_{PS}$ is shown in Fig. \ref{fig1}.
At small temperatures and densities $G_{S}\neq G_{PS}$ signalling a breaking of the chiral symmetry already at the level of the effective Lagrangian. 
The effective quark mass is high at this regime.
Dynamical restoration of the chiral symmetry at large temperatures and densities is manifested by the asymptotic vanishing of the effective quark mass and equality of the scalar and pseudoscalar couplings.

\begin{figure}[t]
\centering
\includegraphics[width=0.49\columnwidth]{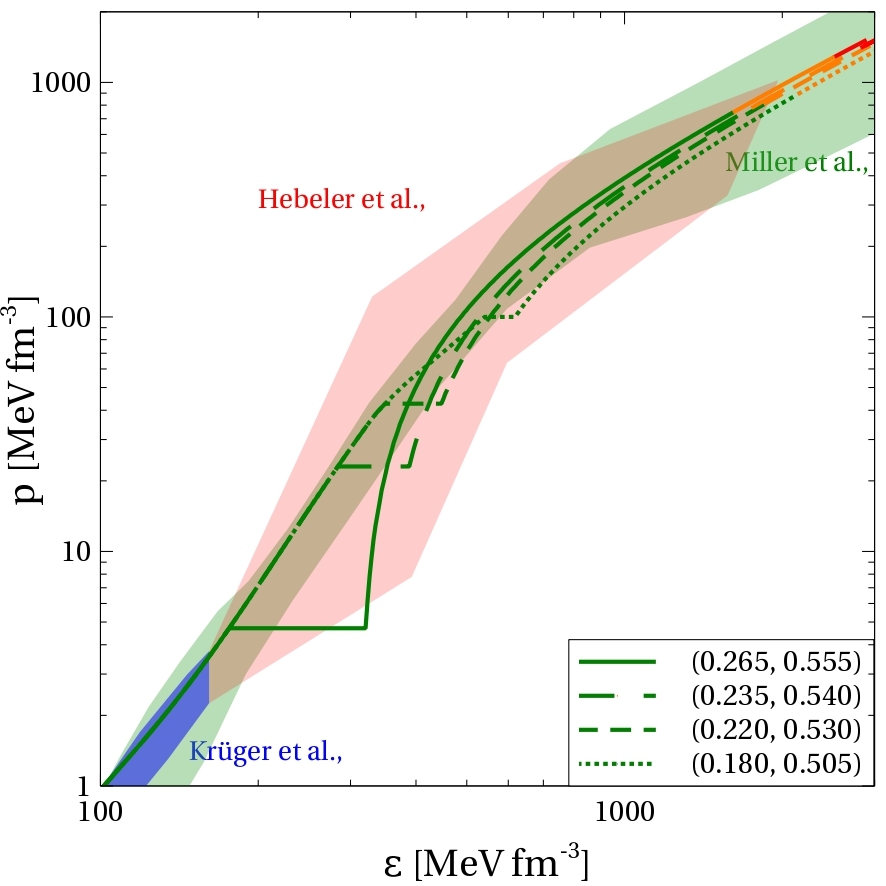}
\includegraphics[width=0.49\columnwidth]{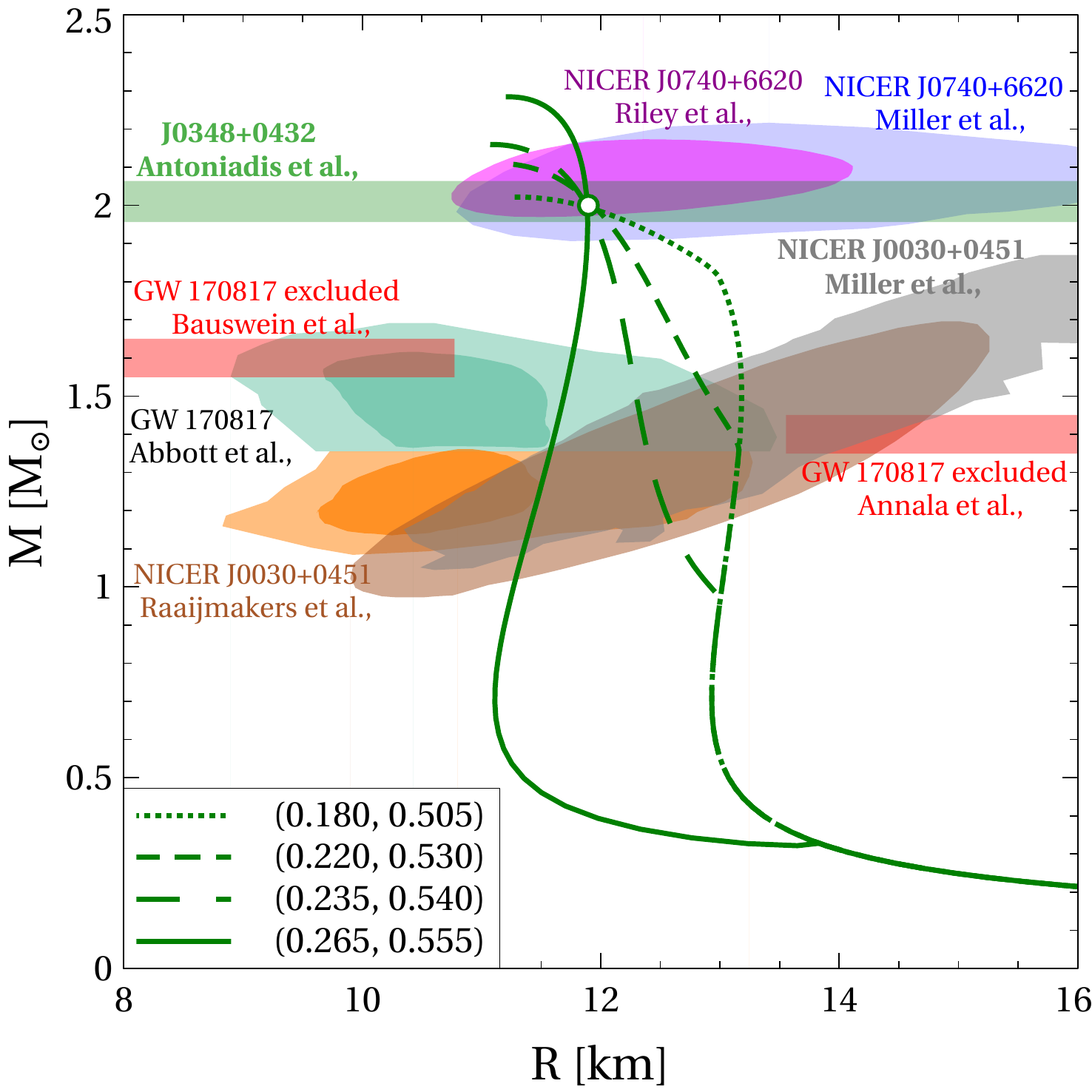}
\caption{Hybrid EoS of cold electrically neutral $\beta$-equilibrated quark-hadron matter in the plane of energy density $\varepsilon$ and pressure
$p$ (left panel) and the corresponding mass-radius relation of NSs (right panel). 
The shaded areas on the left and right panels represent the constraints on the NS EoS \cite{Kruger:2013kua,Hebeler:2013nza,Miller:2021qha} and mass-radius relation \cite{Miller:2021qha,Antoniadis:2013pzd,Riley:2021pdl,Miller:2019cac,Raaijmakers:2019qny,LIGOScientific:2018cki,Bauswein:2017vtn,Annala:2017llu}. 
The green, orange and red parts of the curves are in agreement, in tension and in contradiction with the pQCD constraint from Ref. \cite{Komoltsev:2021jzg}, respectively. The empty circle on the right panel indicates the special point  discussed in the text.}
\label{fig2}
\end{figure}

Having the model parameters fixed we construct the quark matter EoS obtained by applying the mean-field approximation. 
We obey charge neutrality by adding a proper amount of electrons with a chemical potential fulfilling the condition $\mu_e=\mu_d-\mu_u$ for $\beta$-equilibrium.
The hybrid quark-hadron EoS is obtained by matching the quark and hadron ones by means of the Maxwell construction corresponding to a first order phase transition.
For the hadron part we use the DD2npY - T EoS \cite{Shahrbaf:2022upc}, which agrees with the low density constraint from the chiral effective field theory \cite{Kruger:2013kua} and includes nucleonic and hyperonic degrees of freedom.
Fig. \ref{fig2} shows several hybrid EoSs.
As is seen, an increase of the diquark coupling lowers the onset density of quark matter and the radius of light and intermediate mass NSs.
At the same time, higher vector couplings correspond to a stiffer EoS of quark matter, which leads to higher NS maximum mass.
The considered EoSs are consistent with the constraints shown on the left panel of Fig. \ref{fig2}.
Depending on the values of the vector and diquark couplings the constraint obtained by propagating the results of perturbative QCD toward small densities \cite{Komoltsev:2021jzg} is respected by these EoSs up to $\varepsilon=1700 -2100~{\rm MeV~fm^{-3}}$. 
This range of $\varepsilon$ completely covers the energy density values reached in the centers of the heaviest NS provided by a given EoS. 
All the considered EoSs are also consistent with various constraint on the NS mass-radius relation shown on the right panel of Fig. \ref{fig2}.
These EoSs also exhibit the so-called special point (SP), 
which is a narrow region of intersection of hybrid star
sequences. 
It is a generic feature of hybrid quark-hadron EoSs \cite{Cierniak:2020eyh}, while its location is independent on both the hadronic EoS and the phase transition construction and order \cite{Blaschke:2020vuy}.  
For the present model it was first reported in \cite{Ivanytskyi:2022oxv} and analyzed in more details in \cite{SP}.
The constraint on the tidal deformability of a $1.4~{\rm M}_\odot$ NS $\Lambda_{1.4} = 190^{+390}_{-120}$ obtained from the analysis of the gravitational wave signal from the binary neutron star merger GW170817 \cite{LIGOScientific:2018cki} discriminates between the quark-hadron matter EoSs with early and late onset of quark matter.
This constraint is fulfilled only if the quark matter onset is below $1.0~{\rm M}_\odot$.
At $(\eta_{V},\eta_D)=(0.235,0.540)$ and $(\eta_{V},\eta_D)=(0.265,0.555)$ our approach yields $\Lambda_{1.4}=398$ and $\Lambda_{1.4}=326$, respectively.
Thus, the constraint on the NS tidal deformability prefers an early onset of quark matter.

\section{Recovering the conformal limit of quark matter}
\label{sec2}

\begin{figure}[t]
\centering
\includegraphics[width=0.49\columnwidth]{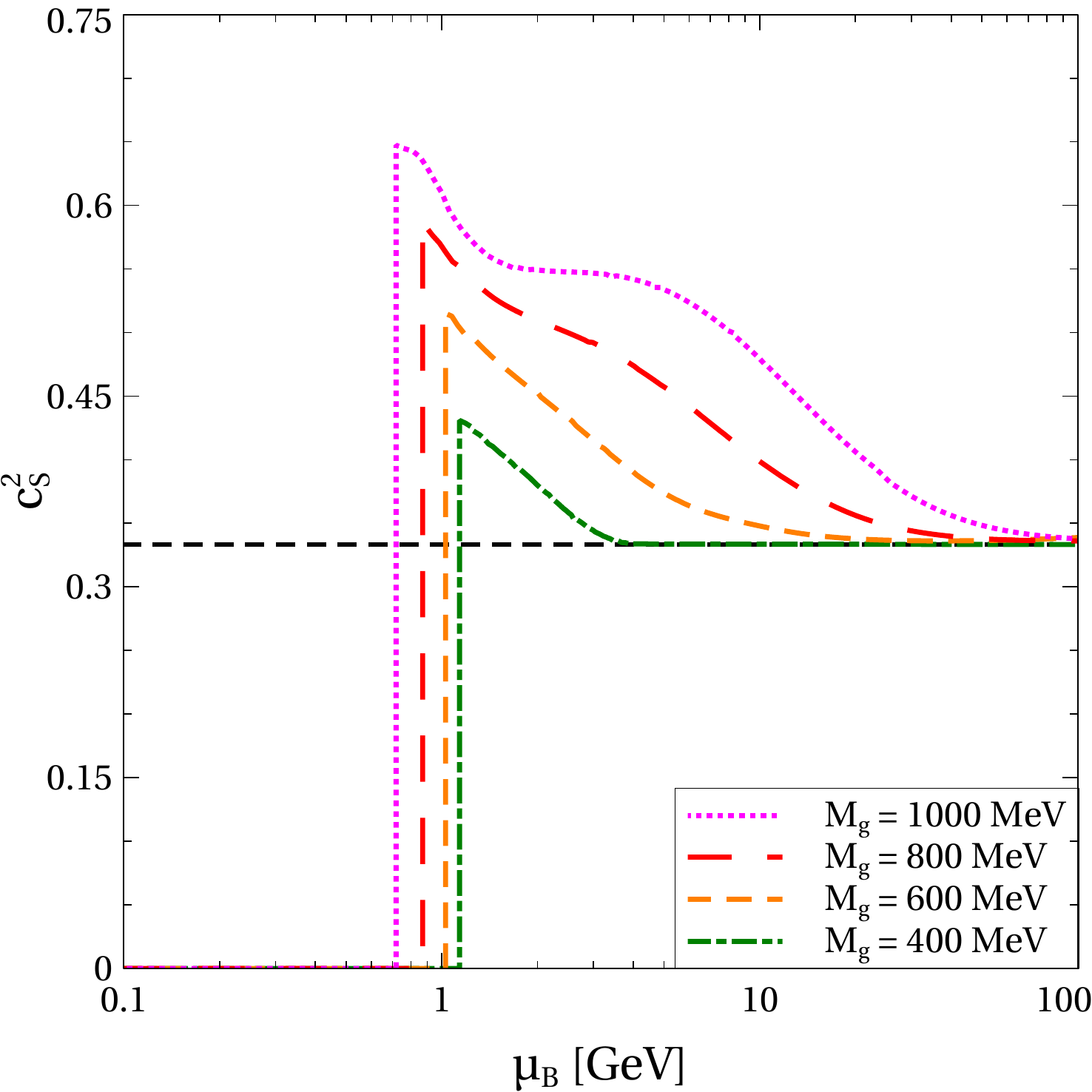}
\includegraphics[width=0.49\columnwidth]{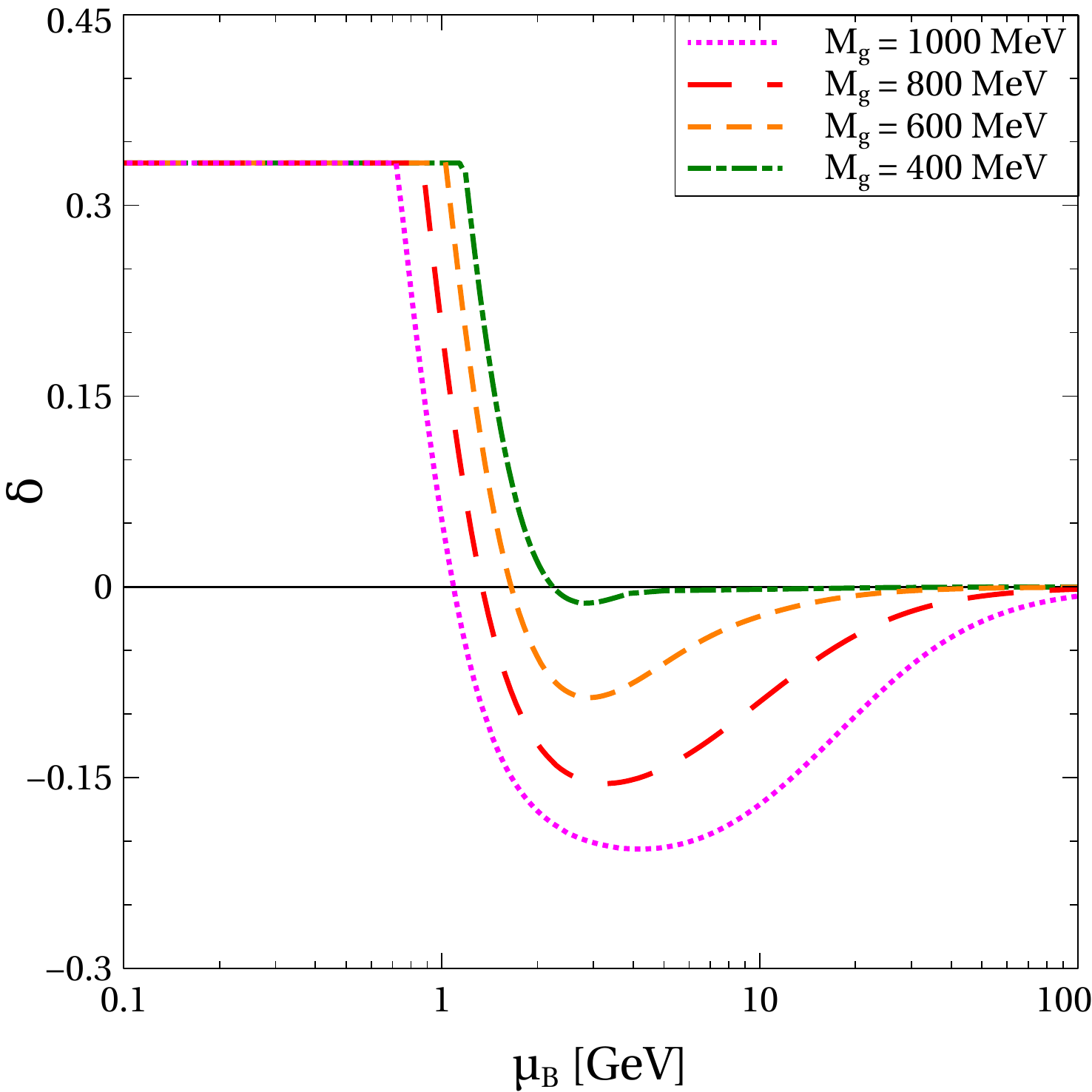}
\caption{Squared speed of sound $c_S^2$ (left panel) and interaction measure $\delta$ (right panel) as functions of baryonic chemical potential $\mu_B$. The black dashed line on the left panel represents $c_S^2=1/3$. Calculations are performed for cold symmetric quark matter at $\eta_V=0.32$ and $\eta_D=0.71$.}
\label{fig3}
\end{figure}

The RDF approach from Ref. \cite{Ivanytskyi:2022oxv} violates the conformal behavior of quark matter expected at high densities. 
This violation is caused by the vector and diquark interaction channels, which are important for the NS phenomenology \cite{Baym:2017whm}. 
Recovering the conformal limit requires suppressing these channels at high densities. 
In Ref. \cite{Ivanytskyi:2022bjc} this was provided by introducing a density dependence of the vector and diquark couplings motivated by exchange of  non-perturbative massive gluons with a mass $M_g$ for QCD in the Landau gauge \cite{Song:2019qoh}. 
Thus,
\begin{eqnarray}
\label{V}
G_V&=&\frac{G_{V0}}{1+\frac{8}{9M_g^2}\left(\frac{\pi^2 \langle q^+q\rangle}{2}\right)^{2/3}},\\
\label{VI}
G_D&=&\frac{G_{D0}}{1+\frac{8}{9M_g^2}\left(\frac{\pi^2 |\langle\overline{q}^ci\tau_2\gamma_5\lambda_2q\rangle|}{2}\right)^{2/3}}.
\end{eqnarray}
Here the vacuum values of the couplings $G_{V0}$ and $G_{D0}$ are parameterized by the dimensionless parameters $\eta_V=G_{V0}/G_{S0}$ and $\eta_D=G_{D0}/G_{S0}$.
Similar to the case of constant vector and diquark couplings these $\eta_V$ and $\eta_D$ are used in order to label EoSs obtained within the present approach.
Eqs. (\ref{V}) and (\ref{VI}) include quark number density $\langle q^+q\rangle$ and diquark condensate $\langle\overline{q}^ci\tau_2\gamma_5\lambda_2q\rangle$, respectively.
They can be found through the partial derivatives of the thermodynamic potential $\Omega$ with respect to quark chemical potentials $\mu_f$ and diquark pairing gap $\Delta$, i.e. $\langle q^+q\rangle=-\sum_f\partial\Omega/\partial\mu_f$ and $|\langle\overline{q}^ci\tau_2\gamma_5\lambda_2q\rangle|=-\partial\Omega/\partial\Delta$. Validity of these expressions is equivalent to thermodynamic consistency of the present approach. 
It is provided by the rearrangement terms 
\begin{eqnarray}
\label{VII}
\Theta_V&=&\int\limits_0^{\langle q^+q\rangle} dn~n^2~\frac{\partial G_V(n)}{\partial n},\\
\label{VIII}
\Theta_D&=&\int\limits_0^{|\langle\overline{q}^ci\tau_2\gamma_5\lambda_2 q\rangle|} dn~n^2~\frac{\partial G_D(n)}{\partial n},
\end{eqnarray}
which contribute $-\Theta_V$ and $\Theta_D$ to the thermodynamic potential. 
At constant couplings these terms vanish identically.
This regime corresponds to the RDF version considered in Section \ref{sec2} and is obtained at $M_g\rightarrow\infty$. 
At finite $M_g$ the present model converges to the conformal limit indicated by $c_S^2\rightarrow 1/3$ and $\delta\rightarrow 0$, where $c_S^2=dp/d\varepsilon$ and $\delta=1/3-p/\varepsilon$ are squared speed of sound and dimensionless interaction measure, respectively. 
As is seen from Fig. \ref{fig3}, smaller values of the non-perturbative gluon mass provide earlier convergence to the conformal regime of the quark matter EoS. The value of this parameter can be estimated based on the Shifman, Vainshtein, and Zakharov expansion of the two-point current correlation functions within massive gauge invariant QCD \cite{Graziani:1984cs}. 
Depending on the value of the frozen QCD fine structure constant and transferred momentum this approach yields 
$M_g=500-1000$ MeV.
The solution of the gluon Schwinger-Dyson equations in the Landau gauge implies $M_g=300-700$ MeV \cite{Cornwall:1981zr,Aguilar:2015bud}.
In this work we use $M_g=600$ MeV as a physically motivated value.

\begin{figure}[t]
\centering
\includegraphics[width=0.49\columnwidth]{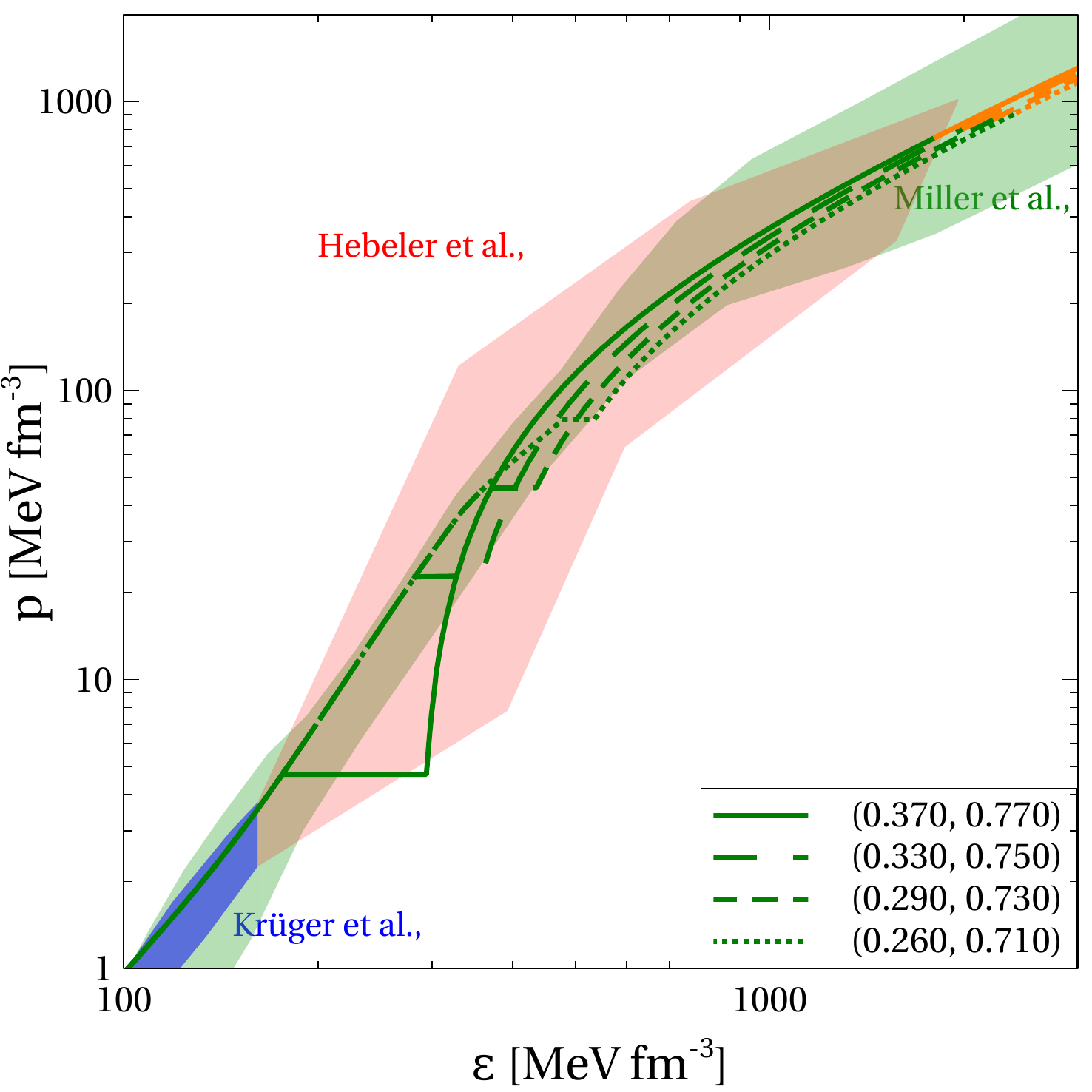}
\includegraphics[width=0.49\columnwidth]{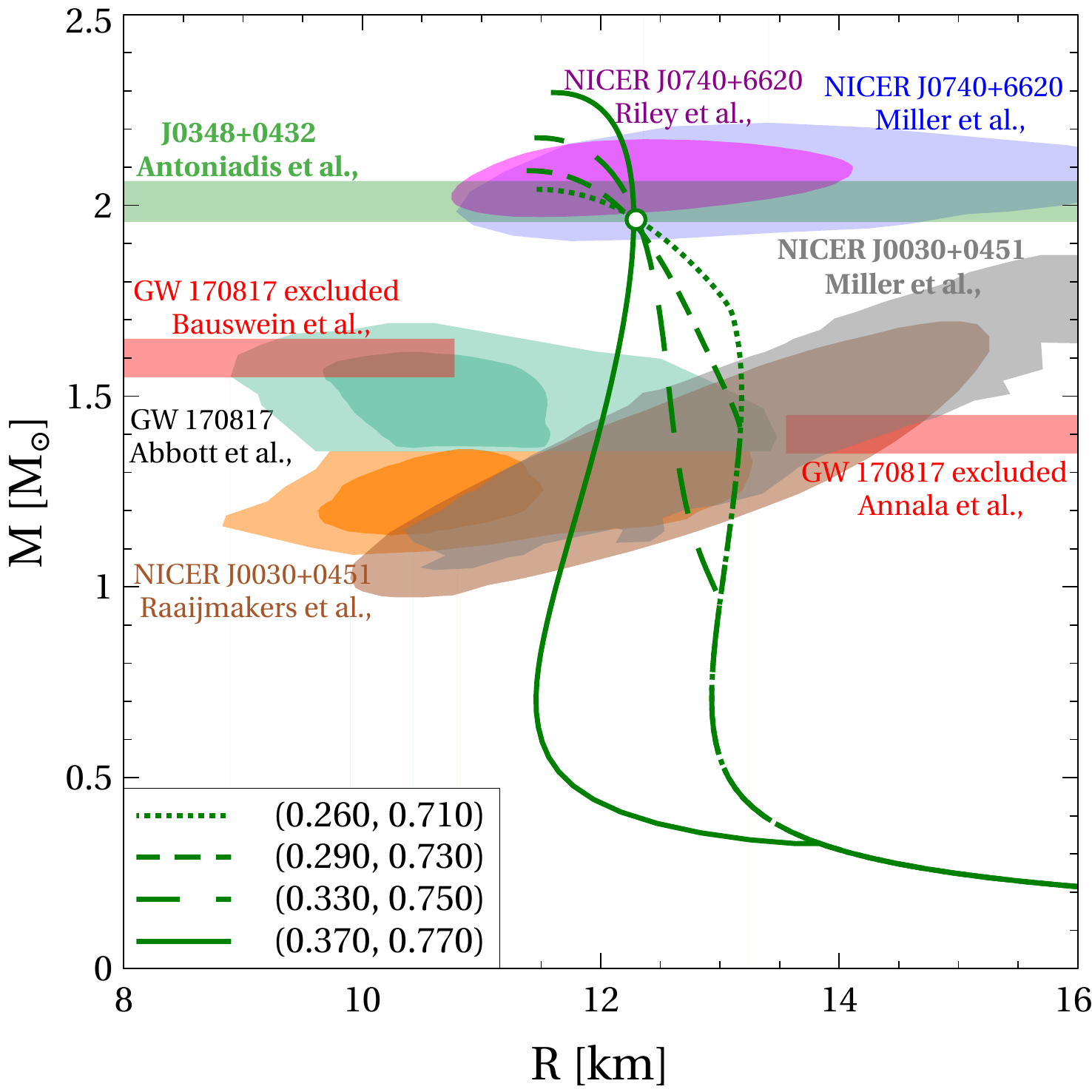}
\caption{The same as in Fig. \ref{fig2} but for the quark matter EoS obtained with density dependent couplings.}
\label{fig4}
\end{figure}

Several EoSs of cold, electrically neutral quark-hadron matter in $\beta$-equilibrium obtained for this value of the non-perturbative gluon mass are shown on Fig. \ref{fig4}.
Similar to the case of constant $G_V$ and $G_D$, these EoSs agree with the constraints mentioned in the previous section. 
At the same time, melting the vector and diquark couplings extends the range of energy densities at which the present approach is consistent with the pQCD constraint \cite{Komoltsev:2021jzg}.
Depending on $\eta_V$ and $\eta_D$ such agreement at any value of the renormalization scale is provided up to $\varepsilon=1800-2200~{\rm MeV~fm}^{-3}$. 
This interval includes all the densities reached in the centers of all stable NSs modelled with the present approach.
The corresponding mass-radius diagrams are shown in Fig. \ref{fig4}.
They are well consistent with the observational constraints mentioned before.
At the same time only the values of $\eta_V$ and $\eta_D$ providing an onset of quark matter below $1.0~{\rm M}_\odot$ are consistent with the constraint on $\Lambda_{1.4}$ from 
Ref. \cite{LIGOScientific:2018cki}.
Similar conclusion was drawn in Ref. \cite{Blaschke:2020vuy}.
For the considered sets of couplings this constraint is respected by $\Lambda_{1.4}=475$ at $(\eta_V,\eta_D)=(0.330,0.750)$ and by $\Lambda_{1.4}=416$ at $(\eta_V,\eta_D)=(0.370,0.770)$.
Finally, similar to the case of constant vector and diquark couplings, the hybrid quark-hadron EoS with density dependent $G_V$ and $G_D$ yields a SP of the mass-radius diagram of NS.

\section{Conclusions}
\label{concl}

We modelled the EoS of color superconducting quark matter within a confining relativistic density functional approach that obeys chirally symmetry of its Lagrangian. 
Its asymptotic convergence to the conformal limit is provided by a melting of the density dependent couplings associated to the vector repulsion and diquark pairing channels. 
The particular dependence of the vector and diquark couplings on the quark number density and diquark condensate, respectively, is motivated by the analysis of the exchange of non-perturbative massive gluon in quark matter.
Matching the developed quark EoS to the hadron one by means of the Maxwell construction allowed us to construct a family of hybrid quark-hadron EoSs which we applied to modelling neutron stars with deconfined quark matter cores.
Applying the recent constraint on the equation of state of dense matter obtained by propagating the results of perturbative quantum chromodynamics toward smaller densities demonstrates that our approach stays reliable in the whole range of densities reached in the interiors of neutron stars.
Furthermore, confronting it to the observational constraints on the mass-radius relation and tidal deformability of neutron stars provides evidence in favor of an early deconfinement of quark matter, which is likely to happen below $1.0~{\rm M}_\odot$.
We also report that the present approach indicates a special point of intersection of hybrid star sequences.

\section*{Acknowledgements}
We acknowledge fruitful discussions throughout the conference, in particular with Alessandro Drago, Oleg Komoltsev, Aleksi Kurkela and Violetta Sagun. 
Consistency with the pQCD constraint was checked using a publicly available Python script by Oleg Komoltsev. 
This work was supported by NCN under grants 2019/33/B/ST9/03059 (O.I., D.B.) and 2020/37/B/ST9/00691 (T.F.). 
A.B. acknowledges support by the European Research Council under the European Union's Horizon 2020 research and innovation program, grant No. 759253, by DFG Project-ID 279384907 - SFB 1245, by DFG - Project-ID 138713538 - SFB 881 and by the State of Hesse within the Cluster Project ELEMENTS. 
The work was performed within a project that has received funding from the Horizon 2020 program under grant agreement STRONG-2020 - No. 824093.



\end{document}